\documentclass[%
 aip,
 amsmath,amssymb,
 reprint,%
]{revtex4-1}

\usepackage{graphicx}
\usepackage{dcolumn}
\usepackage{bm}
\usepackage{multirow}

\usepackage[utf8]{inputenc}
\usepackage[T1]{fontenc}
\usepackage{mathptmx}
\usepackage{color}

\usepackage[page]{appendix}

\usepackage{caption}
\usepackage{subcaption}

\begin{document}

\preprint{AIP/123-QED}

\title[]{
Core Excitations with Excited State Mean Field and Perturbation Theory
}

\author{Scott M. Garner}
\affiliation{
Department of Chemistry, University of California, Berkeley, California 94720, USA 
}
\affiliation{Chemical Sciences Division, Lawrence Berkeley National Laboratory, Berkeley, CA, 94720, USA}

\author{Eric Neuscamman}
 \email{eneuscamman@berkeley.edu.}
\affiliation{
Department of Chemistry, University of California, Berkeley, California 94720, USA 
}
\affiliation{Chemical Sciences Division, Lawrence Berkeley National Laboratory, Berkeley, CA, 94720, USA}

\date{\today}

\begin{abstract}
We test the efficacy of excited state mean field theory and
its excited-state-specific perturbation theory on the prediction
of K-edge positions and X-ray peak separations.
We find that the mean field theory is surprisingly accurate,
even though it contains no accounting of
differential electron correlation effects.
In the perturbation theory, we test multiple
core-valence separation schemes and find that, with the
mean field theory already so accurate, electron-counting
biases in one popular separation scheme become a dominant
error when predicting K-edges.
Happily, these appear to be relatively easy to correct for,
leading to a perturbation theory for K-edge positions
that is lower scaling and more accurate than
coupled cluster theory and competitive in accuracy
with recent high-accuracy results from restricted open-shell
Kohn Sham theory.
For peak separations, our preliminary data show excited state
mean field theory to be exceptionally accurate, but more
extensive testing will be needed to see how it and its
perturbation theory compare to coupled cluster peak separations
more broadly.
\end{abstract}

\maketitle

\section{Introduction}
\label{sec:intro}

X-ray-induced core excitations are probably the most extreme example
of post-excitation orbital relaxations under regular theoretical study.
\cite{norman2018XRayReview}
With core shielding greatly reduced, the valence electrons contract
substantially following the excitation, which means that methods
attempting to predict core excitation energies face a challenge
in getting even the mean-field orbital description right before
even worrying about whether the correlation treatment 
is balanced between ground and excited states.
Unlike charge transfer excitations, where orbital relaxations are
important but ``only'' have a few eVs of energetic impact,
orbital relaxations for core excitations can change energies by 10
eV or more, as shown for example in the difference between
simple configuration interaction singles (CIS) and
non-orthogonal CIS (NOCIS).
\cite{Oosterbaan2018,oosterbaan2020generalized}
From this perspective, it is not too surprising that high-level
correlation treatments like equation of motion (EOM) coupled cluster
\cite{coriani2015,vidal2019new,krylov2020fcCVSEOMDysonOrbitals,Coriani2020CVSCCSD}
and algebraic diagrammatic construction (ADC)
\cite{wenzel2015,fransson2018ADC}
often see errors greater than an eV when predicting the position
of the K-edge.
While these methods have sophisticated correlation treatments,
their ability to relax the orbitals, although present, is limited.
For example, in EOM coupled cluster with singles and doubles (EOM-CCSD),
orbital relaxations come from the doubles part of the configuration
interaction coupling to the primary single excitation,
\cite{Krylov2008}
and so only the first term in the Taylor expansion of a proper
unitary orbital rotation is present.
Given the high cost scaling of high-level correlation treatments
and the fact that their correlation sophistication may be hidden
by incomplete orbital relaxation, it seems worthwhile to explore
methods that work first to fully relax the orbitals and only
then worry about correlation.

The most well-known example of full orbital relaxation is
the $\Delta$SCF family of approaches, \cite{Gilbert2008,Besley2009}
in which an open-shell Slater determinant's orbitals are relaxed
by finding the energy stationary point corresponding to the
desired state.
As Hartree-Fock-based $\Delta$SCF ($\Delta$SCF/HF)
often makes K-edge errors of multiple eV, \cite{Oosterbaan2018}
one would at first glance expect to
get down to errors of an eV or less only when
differential electron correlation is accounted for.
However, things are not necessarily better when
moving to density functional theory (DFT) based
$\Delta$SCF, in which full orbital relaxation
is paired with a state-specific correlation treatment.
Indeed, K-edge errors are still often above an eV and
can be strongly functional dependent.
\cite{Oosterbaan2018}
Very recently, the imposition of approximate spin symmetry
(approximate in that there is no actual wave function to
be spin-symmetric)
via the restricted open-shell Kohn Sham (ROKS) approach
has been shown to offer much greater accuracy in K-edge
prediction for some functionals. \cite{Dip:2020:roks_core}
This success begs a question:  is the improvement due to
the effect of spin-symmetry on the orbital relaxation,
its effect on the correlation treatment, or both?
As we will see below, it would appear that, once spin
symmetry is fully in place, even a correlation-free
mean field treatment of the K-edge becomes accurate
to better than an eV.
Thus, although differential correlation effects
are critical for very high accuracy, it is not clear that
they are the correct explanation for why many
$\Delta$SCF approaches commonly error by multiple eV.

When one does turn to correlation treatments, difficulties
related to valence continuum coupling becomes a key issue.
Very often, quantum chemistry methods separate themselves
from the continuum using the concept of core-valence
separation (CVS), \cite{cederbaum1980cvs}
which has a variety of practical realizations
\cite{peng2019cvsodc,vidal2019new,coriani2015,wenzel2015CVSADCAnalysis,cederbaum1981CVSexcitations,STEX2}
that are not entirely equivalent to each other.
\cite{herbst2020quantifying}
Some ``strong'' CVS approaches disable all core-valence correlation,
\cite{peng2019cvsodc,wenzel2015CVSADCAnalysis,cederbaum1981CVSexcitations,STEX2,vidal2019new}
which is the analogue of the ground state frozen core approximation.
Others \cite{coriani2015,peng2019cvsodc,wenzel2015CVSADCAnalysis}
disable only the correlation terms corresponding to the Auger
processes that are responsible for the actual coupling to
the valence continuum.
For example, the difference between the popular CVS-ADC(2)
and CVS-ADC(2)-x schemes is the treatment of doubly core
excited components of the wave function.
\cite{wenzel2015CVSADCAnalysis}
Although the strong separation approaches are somewhat simpler,
they create a situation in which more electrons are being correlated
in some states as compared to others, an issue that we will
discuss in some detail below.

In this study, we explore the efficacy of excited state
mean field (ESMF) theory
\cite{Shea2018,Shea2020gvp,zhao2020esmf,hardikar2020}
and its excited state analogue (ESMP2)
\cite{Shea2018,Shea2020gvp,Clune2020topesmp2}
to ground state M\o{}ller Plesset theory
for the prediction of core excitation energies.
Like $\Delta$SCF and ROKS, ESMF offers full orbital
relaxation at a cost scaling equivalent to ground
state Hartree Fock theory.
In fact, its practical cost can be as low as a factor
of two when compared to Hartree Fock thanks to the
recent development of a self consistent field
formulation of the theory. \cite{hardikar2020}
Unlike $\Delta$SCF and ROKS, ESMF contains an explicit
ansatz ansatz that is rigorously spin-symmetric,
which, based on the
results presented below, seems to matter
a good deal in allowing it to out-perform
$\Delta$SCF/HF.
As with other post-mean-field correlation treatments,
a CVS approach makes ESMP2 much simpler to apply to
core excitations.
However, there is some question as to what the
ideal CVS approach to ESMP2 is, given that
cancellation of error is key and that most
methods we can compare to do not start from a
fully orbital-relaxed reference, raising the
concern that what makes for good error cancellation
elsewhere may or may not be effective for ESMP2.
As we will see, ESMF is accurate enough on its
own that some choices of CVS for ESMP2 fail 
to make it more accurate than the uncorrelated
reference state!

\section{Theory}
\label{sec:theory}

\subsection{The Single-CSF Reference}
\label{sec:single_csf}

In this study, we restrict ourselves to a simple excited state reference
function in order to simplify our choices for CVS, and because 
good accuracy can be achieved even with this simple choice.
Specifically, our reference will be one singly-excited
singlet-paired configuration state function (CSF) whose
orbitals are optimized state-specifically.
Note that this reference is equivalent to an ESMF wave function whose
configuration interaction coefficients are all zero except for
the up- and down-spin versions of one particular occupied-to-virtual
promotion.
It can also be seen as a two-orbital restricted active space (RAS)
wave function, or as a spin-pure generalization of the
$\Delta$SCF approach.

In practice, preparing this single-CSF reference
for a particular excited state involves selecting an initial
orbital basis (we start with restricted Hartree Fock orbitals), choosing
which occupied-to-virtual transition within this basis will
be used for constructing the initial CSF, and then relaxing the
orbitals so that the energy becomes stationary with respect to
further orbital rotations.
For the basis and states investigated here, we find that the
recently-introduced self consistent field ESMF approach
\cite{hardikar2020}
converges healthily without collapsing
to a different state.
That said, there doubtless are cases where, as in ground state Hartree Fock,
this self consistent field approach will either fail to converge
or converge to wrong stationary point.
In particular, we expect to face convergence problems for higher-lying
states with significant Rydberg character.
In such cases, the quasi-Newton minimization of a generalized
variational principle \cite{Shea2020gvp} can be used
to ensure convergence to the correct state.
Again, in the present study, this was not necessary, allowing
us to benefit from the remarkably higher efficiency \cite{hardikar2020}
of the self consistent field approach.

\begin{figure}[b]
  \includegraphics[width=3.4in,]{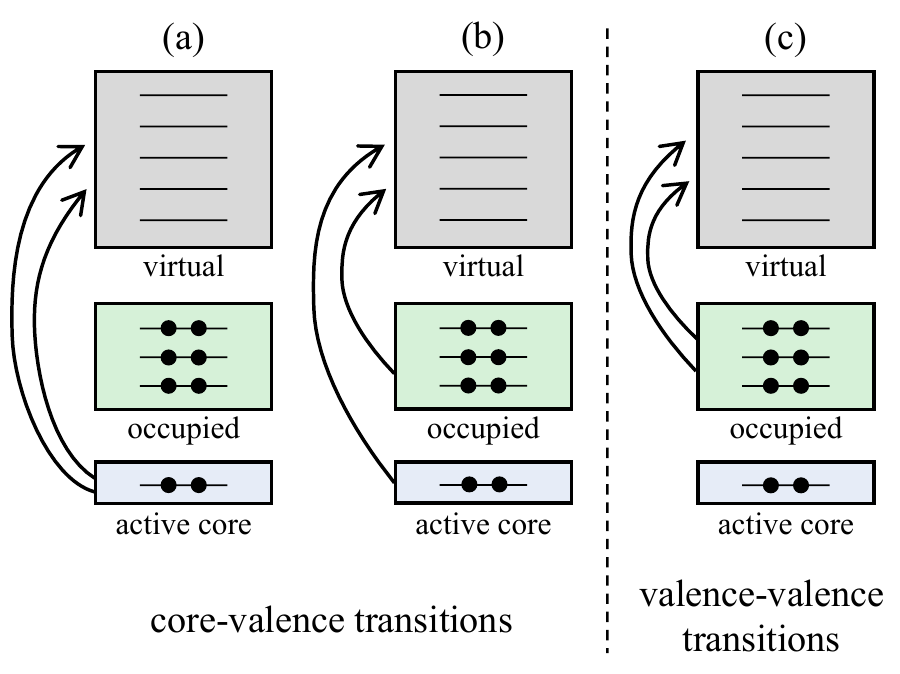}
  \caption{
  Classes of post-RHF excitations that can
  contribute to the MP2 energy.
  The Ao-CVS approach retains all of these terms, while
  the S-CVS approach disables the core-valence terms.
  Note that excitations out of core orbitals that are
  doubly occupied in the reference CSF are neglected in
  both the ground and excited state for S-CVS,
  but included in the ground and excited state for Ao-CVS.
  }
  \label{fig:gs_terms}
\end{figure}

\begin{figure*}[t]
  \includegraphics[width=6.8in,]{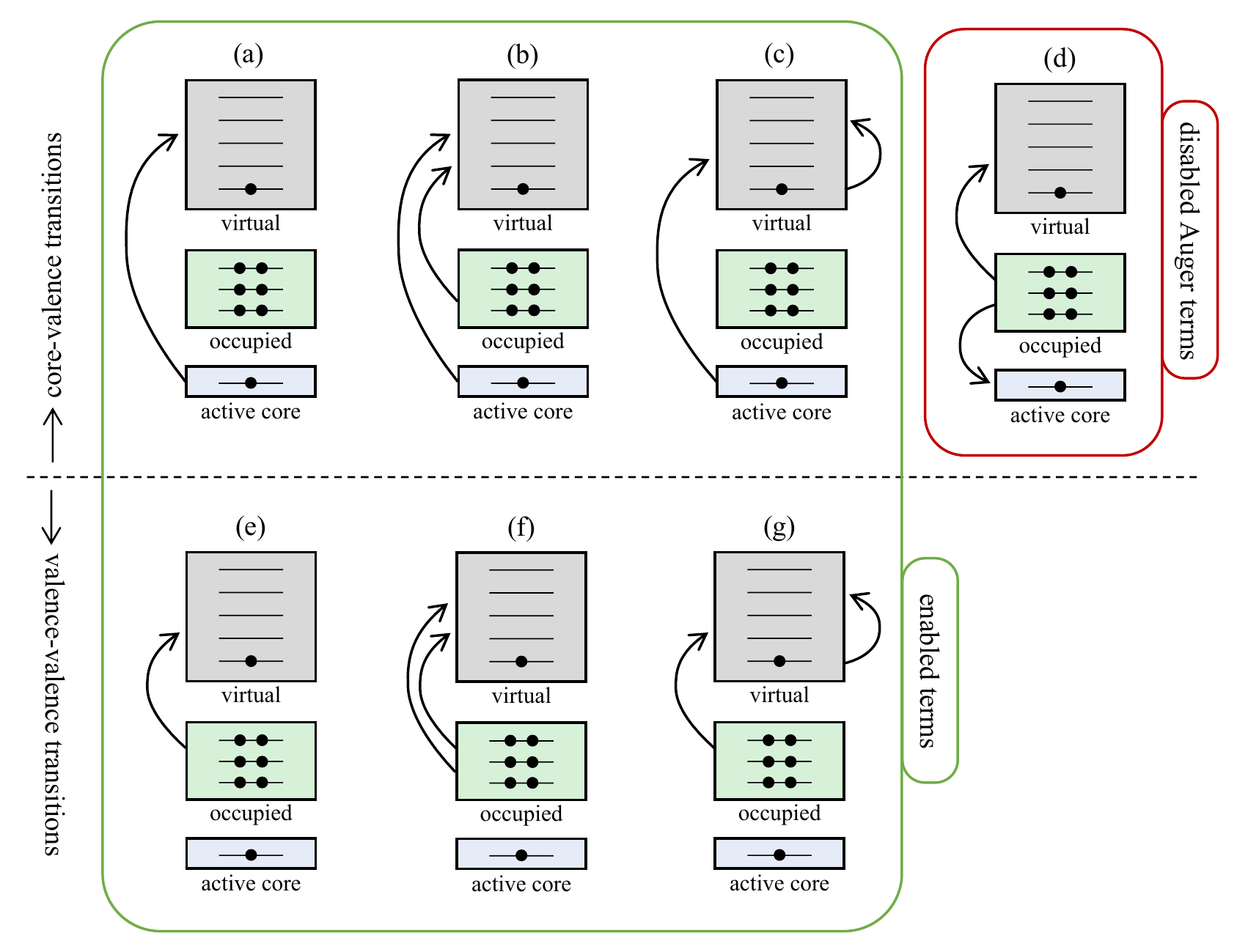}
  \caption{
  Classes of post-ESMF excitations that can
  contribute to the ESMP2 energy when the ESMF
  state is limited to a single core-excited CSF.
  In S-CVS, we retain only the valence-valence terms, and
  core orbitals that are doubly occupied in the reference CSF
  are frozen during the correlation treatment.
  In Ao-CVS, we retain all terms in the ``enabled'' box,
  and core orbitals that are doubly occupied in the reference
  CSF are treated the same as doubly occupied valence orbitals.
  \label{fig:es_terms}
  }
\end{figure*}

\subsection{ESMP2 and the choice of CVS}
\label{sec:cvs_choice}

In this study we will rely on the recently-introduced
$N^5$-scaling version of ESMP2. \cite{Clune2020topesmp2}
To decouple the theory from the valence continuum,
we will test two approaches to
CVS, both of which are closely related to CVS
schemes that have been employed in other theories.
One approach is to disable only out-of-reference
transitions that correspond
to Auger processes, as these terms are often small
in magnitude \cite{coriani2015}
and are the only terms that create numerical
difficulties by coupling to the valence continuum.
For ground state MP2, which we use when evaluating ESMP2
excitation energies, this implies that no orbitals are frozen,
and so this Auger-only-CVS (Ao-CVS) means keeping all
terms in Figure \ref{fig:gs_terms} and all the
terms in the ``enabled'' box of Figure \ref{fig:es_terms}.
This approach is very similar to that used by
Coriani and Koch, \cite{coriani2015}
where the energetic effect of putting the Auger terms
back was typically below 0.1 eV.
These results agree with our finding that, if one enables the
Auger terms, then in the aug-cc-pCVTZ basis at least
ESMP2 can in some cases be converged and gives an energy less
than 0.1 eV different than the more stable Ao-CVS-ESMP2 approach.

A second CVS approach is to disable all out-of-reference
transitions that change the number of electrons in the core.
This ``strong'' CVS (S-CVS) approach is based on the idea
\cite{cederbaum1980cvs}
that the coulomb repulsion integrals that create
these transitions are small due to the small spatial
overlaps between core and valence orbitals.
Setting these integrals to zero is equivalent to disabling
all terms on the core-valence side of the dashed lines in
Figures \ref{fig:gs_terms} and \ref{fig:es_terms}.
Comparing to the recent
Linear-Response Density Cumulant Theory work of
Sokolov \textit{et al}, \cite{peng2019cvsodc}
our S-CVS is similar to their CVS-ODC-12-a scheme in which
their Hessian space is restricted to containing only elements
with exactly one core excitation and up to one additional
valence excitation.
Our Ao-CVS approach, on the other hand, is analogous to
CVS-ODC-12-b in that it allows additional excitations
out of the active edge core.
As we will see in the results, the difference between our
schemes is typically about 1 eV, which is quite similar to
the difference between their two schemes.
Although this type of strong CVS approach has been
used with significant
success for predicting peak separations in the
fc-CVS-EOM-CCSD work of Vidal and coworkers,
\cite{vidal2019new}
we find that it does introduce a noticeable bias
towards lower K-edge energies due to the fact that
the excited state calculations are now correlating
one more electron than the ground state calculations.
Although this bias does not seem to
effect energetic separations between peaks,
there is a relatively simple way to counteract its
effect on K-edge prediction.
As the issue is that we are correlating an extra
electron in the excited state vs the ground state,
we simply need a rough estimate for that extra
correlation energy.
To get this estimate, we should account for
i) the valence electrons in the excited state
experiencing a higher effective nuclear charge
and ii) the fact that the extra electron is
in a singly occupied orbital, and will thus
bring somewhat different correlations with it
than those seen in doubly occupied orbitals.
One approach that accounts for both of these
effects is to evaluate \textit{part} of the
valence correlation energy of the
equivalent-core-approximation \cite{jolly1970,aagren1988}
cation formed by replacing the nucleus in question
by the one with the next higher charge,
e.g.\ C $\rightarrow$ N.
In particular, we sum only the terms in
MP2 in which the occupied indices are distinct
and both in the valence orbitals.
This inter-orbital correlation energy
is then divided by the number of valence electrons
and subtracted from our S-CVS-ESMP2 correlation energy
as a correction to counteract the effect of
having an additional correlated electron relative
to the ground state.
Note that this corrected S-CVS (CS-CVS) approach
has no effect on peak separations, and so if the
plan is to shift the predicted spectrum to line
it up with experiment, then CS-CVS will have
no effect.
Instead, it is intended as a way to improve
predictions of K-edge transition energies.
In future, more sophisticated correlation treatments
may allow for greater insight into how this electron
counting bias is best addressed, but for now
we will rely on the simple after-the-fact correction
discussed above.



\subsection{Artificial Core Hole Localization}
\label{sec:localization}

Having chosen a single-CSF reference, it is important to recognize
that our approach is unable to treat symmetry-delocalized core holes
in a fully rigorous manner.
To take ethylene as an example, the true core excited state contains
by symmetry two components, one with the core hole on the left carbon
and the other with the core hole on the right.
As in any core excited state, the valence orbitals relax strongly
by contracting towards the hole, but as the two components
of the wave function have the hole in different locations, they
experience different relaxation effects.
In other words, the behavior of the valence electrons is strongly
dependent on the positions of the three remaining core electrons,
which is a strong electron
correlation.
Like ground state mean field theory, ESMF theory does not contain
such correlations, and so as a reference it is qualitatively
incorrect for such states.
Note that this is true whether ESMF is used in its full multi-CSF form
or in the limited single-CSF form used here.
Instead, to be qualitatively correct, the reference would need
to build in these strong correlations, as is accomplished for example 
by the NOCIS approach. \cite{Oosterbaan2018}

Although we will show below the poor results that come from blindly
applying ESMP2 to such states, there is a relatively simple
workaround.
As the energy difference between the true hole-delocalized excited
state and the similar broken-symmetry state (in which the hole
is localized on one site) is typically small, \cite{Dip:2020:roks_core}
we can construct a practical approach to predicting these
state's energies by providing our self consistent ESMF optimization
with a hole-localized initial guess.
After identifying the gerade and ungerade 1s Hartree Fock orbitals,
we can prepare an initial guess for our core hole by taking either
the plus or minus combination.
Upon optimizing ESMF, we find in all cases studied here that the SCF
procedure keeps the hole localized, and although this is qualitatively
incorrect, it does not appear to have any adverse effects when
evaluating the ESMP2 correlation energy.
In the ground state calculations for the two CVS schemes,
we note that we are either freezing both
the gerade and ungerade orbitals (S-CVS)
or freezing neither of them (Ao-CVS),
and so we need make no alteration in how the ground state correlation
calculations are done for these states.


\section{Results}
\label{sec:results}

\subsection{Comparing methods}
\label{sec::compare_methods}

Before discussing results for ESMP2 and the two CVS
schemes, let us first consider the accuracy of
our orbital-relaxed ESMF reference wave function on its own.
As seen in Table \ref{tab::kedges}, this single-CSF
ansatz predicts K-edges mostly within
1 eV of experiment, with a mean unsigned error of 0.6 eV.
This result shows that correlation effects cancel in
the K-edge energy difference to a very significant degree.
Remarkably, this mean unsigned error for ESMF
--- which offers no treatment of weak correlation effects ---
is slightly better than what was achieved
by $\Delta$SCF/B3LYP and $\Delta$SCF/MP2,
both of which show mean unsigned errors between
0.6 eV and 0.7 eV for the K-edge values in Table 2 of
Gill et al.\cite{Besley2009}
While one might expect the formal similarities
between ESMF and $\Delta$SCF/HF
to lead to similar overall accuracy,
$\Delta$SCF/HF errors are often
in the 2 to 3 eV range, \cite{Oosterbaan2018}
even when we leave aside molecules with delocalized
core holes.
We therefore see that the slightly more complex
two-determinant ESMF reference is, in this context
at least, well worth the trouble.
Another approach with significant similarity to ESMF
is NOCIS, in which orbital relaxations are approached by
restricted open-shell Hartree Fock calculations on
the relevant cations.
While its K-edge errors are dramatically smaller than
those of orbital-unrelaxed CIS, \cite{Oosterbaan2018}
they are larger than those for ESMF,
which further emphasizes the value of fully state-specific
orbital relaxations.
Of the six functionals tested with ROKS in recent work
by Hait and Head-Gordon, \cite{Dip:2020:roks_core}
only the SCAN and $\omega$B97X-V functionals lead ROKS
to meaningfully better performance than ESMF
in terms of mean unsigned error.
In terms of K-edge energetics, we therefore see that
ESMF's approach of spin-symmetric,
fully-state-specific orbital relaxation coupled with
a complete neglect of weak correlation effects
out-performs NOCIS and $\Delta$SCF/HF
and is comparable to or better than
most approaches that include weak correlation treatments.
One begins to wonder whether the post-ESMF cost of
ESMP2 is going to be worth the trouble.

\begin{table}[t]
\caption{K-Edges (eV) for small molecules.
         The atom in bold is the active edge.
         Estimated relativistic corrections \cite{RelativisticCorrections}
         were added to all calculated excitation energies
         (0.10, 0.21, 0.38, and 0.64 eV for C, N, O, and F, respectively).
         The geometries were set to their experimental values as given
         in the CCCBDB.
         The basis set is aug-cc-pCVTZ.
         \label{tab::kedges}
        }
\begin{tabular}{l c r r r r}
\hline \hline
\multicolumn{3}{c}{\rule{0pt}{3.7mm}} &
\multicolumn{3}{c}{ --------- ESMP2 --------- } \\
Molecule & Exp. & ESMF & S-CVS & CS-CVS & Ao-CVS  \\
\hline

\underline{\textbf{C}}H$_4$ & 288.0\cite{WaterAmmoniaMethane} & 0.5 & -0.7 & -0.2 & 0.3\rule{0pt}{4.2mm}\\
\underline{\textbf{C}}$_2$H$_2$ & 285.8\cite{tronc1979carbon} & 0.7 & -0.8 & -0.1 & 0.3\\
\underline{\textbf{C}}$_2$H$_4$ & 284.7\cite{tronc1979carbon} & 0.7 & -0.8 & -0.1 & 0.3\\
\underline{\textbf{C}}$_2$H$_6$ & 286.9\cite{hitchcock1977carbon} & 0.5 & -0.9 & -0.3 & -0.3\\
H\underline{\textbf{C}}N & 286.4\cite{hitchcock1979inner} & 0.7 & -0.6 & 0.1 & 0.5\\
\underline{\textbf{C}}$_2$N$_2$ & 286.3\cite{hitchcock1979inner} & 0.8 & -0.5 & 0.3 & 0.6  \\
\underline{\textbf{C}}O & 287.4\cite{tronc1979carbon} & 0.8 & -1.1 & -0.2 & 0.0\\
\underline{\textbf{C}}O$_2$ & 290.8\cite{tronc1979carbon} & 1.3 & -1.0 & 0.0 & 0.1\\
H$_2$\underline{\textbf{C}}O & 285.6\cite{remmers1992high} & 0.9 & -0.4 & 0.4 & 0.7\\
\underline{\textbf{C}}H$_3$OH & 287.9\cite{hempelmann1999resonant} & 1.1 & -0.4 & 0.4 & 0.2\\
\underline{\textbf{N}}H$_3$ & 400.7\cite{WaterAmmoniaMethane} & 0.4 & -0.8 & -0.2 & -0.6\\
\underline{\textbf{N}}$_2$ & 400.9\cite{tronc1980nitrogen} & 0.5 & -0.4 & 0.4 & 0.6\\
\underline{\textbf{N}}NO & 401.1\cite{tronc1980nitrogen} & 0.3 & -0.6 & 0.4 & 0.5\\
N\underline{\textbf{N}}O & 404.7\cite{tronc1980nitrogen} & 0.4 & -0.3 & 0.7& 0.7\\
HC\underline{\textbf{N}} & 399.7\cite{hitchcock1979inner} & 0.4 & -0.5 & 0.2 & 0.6\\
C$_2$\underline{\textbf{N}}$_2$ & 398.9\cite{hitchcock1979inner} & 0.7 & -0.5 & 0.3 & 0.6\\
H$_2$\underline{\textbf{O}} & 534.0\cite{WaterAmmoniaMethane} & 0.3 & -0.5 & 0.2 & -0.3\\
C\underline{\textbf{O}} & 534.2\cite{tronc1979carbon}  & -0.2 & -0.4 & 0.2 & 0.7\\
C\underline{\textbf{O}}$_2$ & 535.3\cite{Barrus1979OKedge} & 0.4 & -0.3 & 0.5 & 0.8\\
NN\underline{\textbf{O}} & 535.1\cite{Barrus1979OKedge} & -0.5 & -0.6 & 0.3 & 0.5\\
H$_2$C\underline{\textbf{O}} & 530.8\cite{remmers1992high} & 0.0 & -0.5 & 0.2 & 0.6 \\
CH$_3$\underline{\textbf{O}}H & 534.1\cite{hempelmann1999resonant} & 0.3 & -0.4 & 0.3 & 0.1\\
H\underline{\textbf{F}} & 687.4\cite{hitchcock1981F} & 0.1 & -0.2 & 0.5 & 0.1\\
\underline{\textbf{F}}$_2$ & 682.2\cite{hitchcock1981F} & -0.8 & -0.3 & 0.5 & 0.3 \\[0.5mm]
\hline
\multicolumn{2}{c}{Mean Signed Error} & 0.4 & -0.6 & 0.2 & 0.3\rule{0pt}{3.7mm}\\
\multicolumn{2}{c}{Mean Unsigned Error} & 0.6 & 0.6 & 0.3 & 0.4 \\
\multicolumn{2}{c}{RMS Error} &  0.6  &  0.6  &  0.3  &  0.5  \\[0.5mm]
\hline
\multirow{2}{*}{C} & Signed Error & 0.8 & -0.7 & 0.0 & 0.3\rule{0pt}{3.7mm}\\
 & Unsigned Error & 0.8 & 0.7 & 0.2 & 0.3 \\[0.5mm]
 \hline
\multirow{2}{*}{N} & Signed Error & 0.5 & -0.5 & 0.3 & 0.4 \\
 & Unsigned Error & 0.5 & 0.5 & 0.4 & 0.6 \\[0.5mm]
 \hline
\multirow{2}{*}{O} & Signed Error & 0.1 & -0.4 & 0.3 & 0.4 \\
 & Unsigned Error & 0.3 & 0.4 & 0.3 & 0.5 \\[0.5mm]
 \hline
\multirow{2}{*}{F} & Signed Error & -0.3 & -0.2 & 0.5 & 0.3 \\
 & Unsigned Error & 0.5 & 0.2 & 0.5 & 0.3 \\[0.5mm]
 \hline \hline
\end{tabular}
\end{table}

To find out, let us begin by considering the S-CVS
approach to ESMP2, which tends to underestimate
the K-edge as shown in Table \ref{tab::kedges}.
To understand why, consider the number of electrons
being correlated in the ground and excited state
calculations.
In the ground state, S-CVS is simply the frozen core
approximation, and so the number of correlated electrons
is equal to the number of valence electrons.
In S-CVS-ESMP2, however, there is an additional
valence electron, so although the core is still
frozen, the number of electrons being correlated
is one larger than in the ground state.
If one considers that an electron pair's correlation
energy is on the order of 1 eV, and that one electron
is half of a pair, a simple electron counting
argument seems to explain the underestimation.
Our CS-CVS approach, which attempts to correct for
this bias in a molecule-specific manner by parsing
MP2 contributions in the one-higher-core-charge cation, 
does manage to improve the K-edge predictions overall,
but it is clearly more effective for the C K-edge
than for  the others.

\begin{table}[t]
\caption{Peak separations (eV) in methane, ammonia, and water.
         Experimental numbers are the peak separations
         between the lowest-lying core excited state
         and higher-lying states,
         with methane's lowest core excited state adjusted to
         remove the $\nu_4$ vibrational quanta. \cite{schirmer1993}
         Theoretical numbers are errors relative to the
         experimental peak separations, with S-CVS
         and CS-CVS (as they are identical for separations)
         reported together as (C)S-CVS.
         \label{tab::peak_sep}
        }
\begin{tabular}{l l r@{.}l r@{.}l r@{.}l r@{.}l}
\hline \hline
\multicolumn{6}{c}{\rule{0pt}{3.7mm}} &
\multicolumn{4}{c}{ ------- ESMP2 ------- } \\
\multicolumn{2}{c}{Transition\rule{0pt}{3.7mm}} &
\multicolumn{2}{c}{Exp. \hphantom{ }} &
\multicolumn{2}{c}{ESMF} &
\multicolumn{2}{c}{(C)S-CVS} &
\multicolumn{2}{c}{Ao-CVS} \\[0.5mm]
\hline 
CH$_4$ & 3$a_1$/3s $\rightarrow$ 2$t_2$/3p & 1&30 &  0&00 & \hspace{3mm} 0&19 & \hspace{3mm} 0&90 \rule{0pt}{3.7mm}\\[0.4mm]
NH$_3$ & 4$a_1$/3s $\rightarrow$ 2$e$/3p   & 1&67 & -0&01 & \hspace{3mm} 0&18 & \hspace{3mm} 1&07 \\[0.4mm]
       & 4$a_1$/3s $\rightarrow$ 5$a_1$/3p & 2&20 &  0&51 & \hspace{3mm} 0&78 & \hspace{3mm} 1&65 \\[0.4mm]
OH$_2$ & 4$a_1$/3s $\rightarrow$ 2$b_2$/3p & 1&89 & -0&01 & \hspace{3mm} 0&10 & \hspace{3mm} 1&06 \\[1.0mm]
\hline
\end{tabular}
\end{table}



Turning now to Ao-CVS, we see in Table \ref{tab::kedges}
that it's K-edges are typically more accurate
than those of S-CVS, but not as accurate as
those of CS-CVS.
Note that, as Ao-CVS correlates all electrons in
both the ground and the excited state, there are no
correlation adjustments to be made based on counting
arguments.
While the K-edges suggest that Ao-CVS and CS-CVS
do have something to offer over ESMF, the
peak separations shown in Table \ref{tab::peak_sep}
are a different story.
Once one accounts for the fact that
the half-eV error for ammonia's 5$a_1$/3p state
is almost entirely due to the fact that
our basis lacks extra Rydberg functions,
\cite{vidal2019new}
ESMF is more accurate for this small initial
sample of peak separations than any of the 
ESMP2 CVS variants.
Indeed, it is, in this Rydberg-deficient basis,
on par with recent equation of motion coupled
cluster work, \cite{coriani2015,vidal2019new}
which is quite remarkable given
its complete neglect of the correlation details.
Ao-CVS-ESMP2, on the other hand, is especially poor
for these peak separations, biasing the energies
of the higher states up by about 1 eV,
which calls for some analysis.
In looking at the various contributions to the Ao-CVS-ESMP2
correlation energy, we find that a key difference between the
(lower) totally symmetric s states and the the (higher)
not-totally-symmetric p states in Table \ref{tab::peak_sep}
is found in the
ESMP2 energy contribution from the
the determinant in which the second core electron
has been promoted to join its partner in the
reference CSF's singly-occupied virtual orbital.
This determinant is totally symmetric in these molecules,
and so cannot make any energy contribution to the
p states.
In the s states, however, this Figure-\ref{fig:es_terms}-type-(a)
determinant contributes about 1 eV of correlation energy,
which accounts for the 1 eV bias against the p states
that we see but does not quite explain it.
The answer may lie in the excitations
that this N$^5$-scaling version of ESMP2 leaves out.
\cite{Clune2020topesmp2}
Many of these neglected excitations are core-valence
in nature, and so the present implementation of
Ao-CVS-ESMP2 may be biased by its lack of these
core-valence terms.
However, in our experience with valence excitations
at least, we have yet to see these terms have any
appreciable effect. \cite{Clune2020topesmp2}
The S-CVS and CS-CVS approaches, in contrast,
neglect all core-valence terms, which according
to the current results is substantially less
biased when evaluating peak separations.
In sum, the CS-CVS approach appears preferrable
to ESMF for evaluating the K-edge position, but
it is not obvious that any version of ESMP2
is preferrable to ESMF when evaluating
peak separations, especially considering
that ESMF can now be run for a near-Hartree-Fock cost.
\cite{hardikar2020}

As a final note on accuracy before we turn our attention
to periodic trends, Table \ref{tab:localization} shows
that both ESMF and ESMP2 do indeed perform quite poorly
if the core orbital is delocalized.
Again, the reason for this failure is that the mean
field reference does not contain the strong correlations
inherent to how the valance electron's contractions are
dependent on the positions of the remaining core electrons.
While one can cheat by artificially localizing the core
hole without
much effect on accuracies in the molecules studied
here, this formal failure of the theory may well cause
more serious problems in other delocalize-core-hole situations.
Certainly the electron density it predicts will be
biased towards one side and will fail to respect symmetry,
which will for example create a permanent dipole moment where
one should not exist.

\subsection{Periodic trends}
\label{sec::trends}

Another interesting aspect of the K-edge results
of Table \ref{tab::kedges} is the periodic trends
they contain.
All else being equal, the per-electron correlation
energy for valence electrons rises as one moves
from lighter (C) to heavier (F) elements, which
one can verify by simple frozen core MP2 calculations
on CH$_4$, NH$_3$, OH$_2$, and FH.
As the valence electrons in a core excited state
experience an effective nuclear charge that is
one higher than in the ground state, the natural
scale of valence correlation energy is larger in the K-edge state.
With ESMF neglecting all correlation outside the (sub-0.1eV)
spin recoupling energy, we would therefore expect it
to error high for K-edges, which is exactly what we see.
Further, as the relative effective change in core charge
is larger in lighter atoms (e.g.\ C$\rightarrow$N grows
the formal charge by 7/6 while F$\rightarrow$Ne grows
it by only 10/9) we would expect ESMF's
tendency to error high to be stronger in lighter atoms,
as these experience a larger relative jump in valence
correlation energy scale upon core excitation.
Again, this is what we see.

Another clear trend is that S-CVS's expected under-estimation
of K-edges gets smaller as the K-edge element gets heavier.
As with ESMF's tendency to over-estimate, one can try to
rationalize this trend based on lighter elements having
larger jumps in the valence correlation energy scale
upon K-edge excitation.
As S-CVS is correlating an extra electron compared to
the ground state, and since this correlation is happening
on the excited state's (larger) correlation energy scale,
one might argue that the fact that the excitation-induced
jump in this scale is larger in light elements explains
why S-CVS has a larger negative K-edge bias in lighter
elements.
However, the fact that the correction we make in
CS-CVS has essentially the same scale (0.7 to 0.8 eV on average)
for the different elements' K-edges suggests that there
may be a second effect contributing to the trend in
the S-CVS K-edge energy biases.

\begin{table}[t]
\centering    
\caption{Comparison of localized- and delocalized-hole approaches
         for the acetylene K-edge in the aug-cc-pCVTZ basis.
         The hole is either delocalized in the gerade (g) or
         ungarade (u) 1s orbital, or localized on left (L) or right (R).
         The experimental edge lies at 285.8 eV. \cite{tronc1979carbon}
         \label{tab:localization}
        }
\begin{tabular}{ l c c c}
\hline \hline
    & ESMF & S-CVS & Ao-CVS \\
\hline
    L & 286.4 & 285.0 & 286.0\\
    R & 286.4 & 285.0 & 286.0\\
    g & 294.1 & 292.5 & 293.0\\
    u & 294.0 & 292.4 & 293.0\\
    \hline \hline
\end{tabular}
\end{table}

\section{Conclusion}
\label{sec:conclusion}

We have investigated the ability of excited state mean
field theory and an accompanying excited-state-specific
perturbation theory
to predict K-edge energies and peak separations
for core excited states.
Our most remarkable finding is that, despite its
blanket neglect of correlation energy,
excited state mean field theory is typically within
1 eV for K-edge energies and even more accurate
in a preliminary test on peak separations.
Indeed, for the latter, it is more accurate than
the associated perturbation theory and competitive
with equation of motion coupled cluster,
emphasizing how large a role correlation energy
cancellation plays in these predictions.
By correcting the electron-counting bias inherent
to one of our perturbation theory's core-valence separation
schemes, we find that the perturbation theory can
out-perform the mean field theory for predicting
the position of the K-edge, and indeed appears to
outperform most other available methods in this regard.
The only available method that seems to do significantly
better is restricted-open-shell Kohn Sham theory, and
then only when the SCAN functional is employed.
In cases where self-interaction errors in the
valence orbitals are a concern (e.g.\ in a
molecule where the LUMO and LUMO+1 are spatially
separate and similar but not degenerate in energy),
the perturbation theory presented here may be
clearly preferrable, especially given that its
$N^5$ cost scaling is lower than the $N^6$
scaling of coupled cluster theory.

Looking forward, there are multiple opportunities and
priorities for further development.
For starters, the after-the-fact correction we use
to counteract our core-valence separation's
electron counting bias is clearly not unique.
In future, it could be more satisfying and predictive
to perform a more in-depth analysis of the different
ESMP2 contributions to the excited state energy.
It may for example be possible to identify and explicitly disable
terms that correspond to the electron counting bias.
In excited state mean field theory, a more efficient implementation
of the quasi-Newton minimization of generalized variational
principles would be especially helpful for states with
Rydberg character, as the self-consistent field approach,
although fast, is often not stable once extended Rydberg
functions are added to the basis.
Thinking of perturbation theory's wide existing role in
supporting other methods, it may also be interesting
to develop ESMP2 natural orbitals as a starting
point for local correlation methods and to test the
efficacy of ESMP2 as a black-box (i.e.\ no active space needed)
generator of multi-Slater expansions in core-excited
quantum Monte Carlo.
Finally, given the efficacy of restricted open-shell Kohn
Sham and the remarkable difference in accuracy between
Hartree-Fock-based $\Delta$SCF and our excited state
mean field theory, it would be especially interesting
to investigate whether density functional extensions \cite{Zhao2019dft}
to this mean field theory can combine the best
of both worlds.

$\vspace{1mm}$

\noindent
\textit{Acknowledgments} ---
This work was supported by the
Office of Science, Office of Basic Energy Sciences,
the U.S.\ Department of Energy,
Contract No.\ {DE-AC02-05CH11231}.
Calculations were performed using
the LBNL Lawrencium computing cluster.

$\vspace{1mm}$

\noindent
\textit{Data Availability Statement} ---
The data that supports the findings of this study
are available within the article.

\bibliography{main}

\begin{thebibliography}{38}%
\makeatletter
\providecommand \@ifxundefined [1]{%
 \@ifx{#1\undefined}
}%
\providecommand \@ifnum [1]{%
 \ifnum #1\expandafter \@firstoftwo
 \else \expandafter \@secondoftwo
 \fi
}%
\providecommand \@ifx [1]{%
 \ifx #1\expandafter \@firstoftwo
 \else \expandafter \@secondoftwo
 \fi
}%
\providecommand \natexlab [1]{#1}%
\providecommand \enquote  [1]{``#1''}%
\providecommand \bibnamefont  [1]{#1}%
\providecommand \bibfnamefont [1]{#1}%
\providecommand \citenamefont [1]{#1}%
\providecommand \href@noop [0]{\@secondoftwo}%
\providecommand \href [0]{\begingroup \@sanitize@url \@href}%
\providecommand \@href[1]{\@@startlink{#1}\@@href}%
\providecommand \@@href[1]{\endgroup#1\@@endlink}%
\providecommand \@sanitize@url [0]{\catcode `\\12\catcode `\$12\catcode
  `\&12\catcode `\#12\catcode `\^12\catcode `\_12\catcode `\%12\relax}%
\providecommand \@@startlink[1]{}%
\providecommand \@@endlink[0]{}%
\providecommand \url  [0]{\begingroup\@sanitize@url \@url }%
\providecommand \@url [1]{\endgroup\@href {#1}{\urlprefix }}%
\providecommand \urlprefix  [0]{URL }%
\providecommand \Eprint [0]{\href }%
\providecommand \doibase [0]{http://dx.doi.org/}%
\providecommand \selectlanguage [0]{\@gobble}%
\providecommand \bibinfo  [0]{\@secondoftwo}%
\providecommand \bibfield  [0]{\@secondoftwo}%
\providecommand \translation [1]{[#1]}%
\providecommand \BibitemOpen [0]{}%
\providecommand \bibitemStop [0]{}%
\providecommand \bibitemNoStop [0]{.\EOS\space}%
\providecommand \EOS [0]{\spacefactor3000\relax}%
\providecommand \BibitemShut  [1]{\csname bibitem#1\endcsname}%
\let\auto@bib@innerbib\@empty
\bibitem [{\citenamefont {Norman}\ and\ \citenamefont
  {Dreuw}(2018)}]{norman2018XRayReview}%
  \BibitemOpen
  \bibfield  {author} {\bibinfo {author} {\bibfnamefont {P.}~\bibnamefont
  {Norman}}\ and\ \bibinfo {author} {\bibfnamefont {A.}~\bibnamefont {Dreuw}},\
  }\bibfield  {title} {\enquote {\bibinfo {title} {Simulating x-ray
  spectroscopies and calculating core-excited states of molecules},}\
  }\href@noop {} {\bibfield  {journal} {\bibinfo  {journal} {Chemical reviews}\
  }\textbf {\bibinfo {volume} {118}},\ \bibinfo {pages} {7208--7248} (\bibinfo
  {year} {2018})}\BibitemShut {NoStop}%
\bibitem [{\citenamefont {Oosterbaan}, \citenamefont {White},\ and\
  \citenamefont {Head-Gordon}(2018)}]{Oosterbaan2018}%
  \BibitemOpen
  \bibfield  {author} {\bibinfo {author} {\bibfnamefont {K.~J.}\ \bibnamefont
  {Oosterbaan}}, \bibinfo {author} {\bibfnamefont {A.~F.}\ \bibnamefont
  {White}}, \ and\ \bibinfo {author} {\bibfnamefont {M.}~\bibnamefont
  {Head-Gordon}},\ }\bibfield  {title} {\enquote {\bibinfo {title}
  {Non-orthogonal configuration interaction with single substitutions for the
  calculation of core-excited states},}\ }\href@noop {} {\bibfield  {journal}
  {\bibinfo  {journal} {J. Chem. Phys.}\ }\textbf {\bibinfo {volume} {149}},\
  \bibinfo {pages} {044116} (\bibinfo {year} {2018})}\BibitemShut {NoStop}%
\bibitem [{\citenamefont {Oosterbaan}\ \emph {et~al.}(2020)\citenamefont
  {Oosterbaan}, \citenamefont {White}, \citenamefont {Hait},\ and\
  \citenamefont {Head-Gordon}}]{oosterbaan2020generalized}%
  \BibitemOpen
  \bibfield  {author} {\bibinfo {author} {\bibfnamefont {K.~J.}\ \bibnamefont
  {Oosterbaan}}, \bibinfo {author} {\bibfnamefont {A.~F.}\ \bibnamefont
  {White}}, \bibinfo {author} {\bibfnamefont {D.}~\bibnamefont {Hait}}, \ and\
  \bibinfo {author} {\bibfnamefont {M.}~\bibnamefont {Head-Gordon}},\
  }\bibfield  {title} {\enquote {\bibinfo {title} {Generalized single
  excitation configuration interaction: an investigation into the impact of the
  inclusion of non-orthogonality on the calculation of core-excited states},}\
  }\href@noop {} {\bibfield  {journal} {\bibinfo  {journal} {Physical Chemistry
  Chemical Physics}\ }\textbf {\bibinfo {volume} {22}},\ \bibinfo {pages}
  {8182--8192} (\bibinfo {year} {2020})}\BibitemShut {NoStop}%
\bibitem [{\citenamefont {Coriani}\ and\ \citenamefont
  {Koch}(2015)}]{coriani2015}%
  \BibitemOpen
  \bibfield  {author} {\bibinfo {author} {\bibfnamefont {S.}~\bibnamefont
  {Coriani}}\ and\ \bibinfo {author} {\bibfnamefont {H.}~\bibnamefont {Koch}},\
  }\bibfield  {title} {\enquote {\bibinfo {title} {Communication: X-ray
  absorption spectra and core-ionization potentials within a core-valence
  separated coupled cluster framework},}\ }\href@noop {} {\bibfield  {journal}
  {\bibinfo  {journal} {J. Chem. Phys.}\ }\textbf {\bibinfo {volume} {143}},\
  \bibinfo {pages} {181103} (\bibinfo {year} {2015})}\BibitemShut {NoStop}%
\bibitem [{\citenamefont {Vidal}\ \emph {et~al.}(2019)\citenamefont {Vidal},
  \citenamefont {Feng}, \citenamefont {Epifanovsky}, \citenamefont {Krylov},\
  and\ \citenamefont {Coriani}}]{vidal2019new}%
  \BibitemOpen
  \bibfield  {author} {\bibinfo {author} {\bibfnamefont {M.~L.}\ \bibnamefont
  {Vidal}}, \bibinfo {author} {\bibfnamefont {X.}~\bibnamefont {Feng}},
  \bibinfo {author} {\bibfnamefont {E.}~\bibnamefont {Epifanovsky}}, \bibinfo
  {author} {\bibfnamefont {A.~I.}\ \bibnamefont {Krylov}}, \ and\ \bibinfo
  {author} {\bibfnamefont {S.}~\bibnamefont {Coriani}},\ }\bibfield  {title}
  {\enquote {\bibinfo {title} {New and efficient equation-of-motion
  coupled-cluster framework for core-excited and core-ionized states},}\
  }\href@noop {} {\bibfield  {journal} {\bibinfo  {journal} {J. Chem. Theory
  Comput.}\ }\textbf {\bibinfo {volume} {15}},\ \bibinfo {pages} {3117--3133}
  (\bibinfo {year} {2019})}\BibitemShut {NoStop}%
\bibitem [{\citenamefont {Vidal}, \citenamefont {Krylov},\ and\ \citenamefont
  {Coriani}(2020)}]{krylov2020fcCVSEOMDysonOrbitals}%
  \BibitemOpen
  \bibfield  {author} {\bibinfo {author} {\bibfnamefont {M.~L.}\ \bibnamefont
  {Vidal}}, \bibinfo {author} {\bibfnamefont {A.~I.}\ \bibnamefont {Krylov}}, \
  and\ \bibinfo {author} {\bibfnamefont {S.}~\bibnamefont {Coriani}},\
  }\bibfield  {title} {\enquote {\bibinfo {title} {Dyson orbitals within the
  fc-cvs-eom-ccsd framework: theory and application to x-ray photoelectron
  spectroscopy of ground and excited states},}\ }\href@noop {} {\bibfield
  {journal} {\bibinfo  {journal} {Physical Chemistry Chemical Physics}\
  }\textbf {\bibinfo {volume} {22}},\ \bibinfo {pages} {2693--2703} (\bibinfo
  {year} {2020})}\BibitemShut {NoStop}%
\bibitem [{\citenamefont {Faber}\ and\ \citenamefont
  {Coriani}(2020)}]{Coriani2020CVSCCSD}%
  \BibitemOpen
  \bibfield  {author} {\bibinfo {author} {\bibfnamefont {R.}~\bibnamefont
  {Faber}}\ and\ \bibinfo {author} {\bibfnamefont {S.}~\bibnamefont
  {Coriani}},\ }\bibfield  {title} {\enquote {\bibinfo {title}
  {Core--valence-separated coupled-cluster-singles-and-doubles
  complex-polarization-propagator approach to x-ray spectroscopies},}\
  }\href@noop {} {\bibfield  {journal} {\bibinfo  {journal} {Physical Chemistry
  Chemical Physics}\ }\textbf {\bibinfo {volume} {22}},\ \bibinfo {pages}
  {2642--2647} (\bibinfo {year} {2020})}\BibitemShut {NoStop}%
\bibitem [{\citenamefont {Wenzel}\ \emph
  {et~al.}(2015{\natexlab{a}})\citenamefont {Wenzel}, \citenamefont {Holzer},
  \citenamefont {Wormit},\ and\ \citenamefont {Dreuw}}]{wenzel2015}%
  \BibitemOpen
  \bibfield  {author} {\bibinfo {author} {\bibfnamefont {J.}~\bibnamefont
  {Wenzel}}, \bibinfo {author} {\bibfnamefont {A.}~\bibnamefont {Holzer}},
  \bibinfo {author} {\bibfnamefont {M.}~\bibnamefont {Wormit}}, \ and\ \bibinfo
  {author} {\bibfnamefont {A.}~\bibnamefont {Dreuw}},\ }\bibfield  {title}
  {\enquote {\bibinfo {title} {Analysis and comparison of cvs-adc approaches up
  to third order for the calculation of core-excited states},}\ }\href@noop {}
  {\bibfield  {journal} {\bibinfo  {journal} {J. Chem. Phys.}\ }\textbf
  {\bibinfo {volume} {142}},\ \bibinfo {pages} {214104} (\bibinfo {year}
  {2015}{\natexlab{a}})}\BibitemShut {NoStop}%
\bibitem [{\citenamefont {Fransson}\ and\ \citenamefont
  {Dreuw}(2018)}]{fransson2018ADC}%
  \BibitemOpen
  \bibfield  {author} {\bibinfo {author} {\bibfnamefont {T.}~\bibnamefont
  {Fransson}}\ and\ \bibinfo {author} {\bibfnamefont {A.}~\bibnamefont
  {Dreuw}},\ }\bibfield  {title} {\enquote {\bibinfo {title} {Simulating x-ray
  emission spectroscopy with algebraic diagrammatic construction schemes for
  the polarization propagator},}\ }\href@noop {} {\bibfield  {journal}
  {\bibinfo  {journal} {Journal of chemical theory and computation}\ }\textbf
  {\bibinfo {volume} {15}},\ \bibinfo {pages} {546--556} (\bibinfo {year}
  {2018})}\BibitemShut {NoStop}%
\bibitem [{\citenamefont {Krylov}(2008)}]{Krylov2008}%
  \BibitemOpen
  \bibfield  {author} {\bibinfo {author} {\bibfnamefont {A.~I.}\ \bibnamefont
  {Krylov}},\ }\bibfield  {title} {\enquote {\bibinfo {title}
  {{Equation-of-motion coupled-cluster methods for open-shell and
  electronically excited species: the Hitchhiker's guide to Fock space.}}}\
  }\href {\doibase 10.1146/annurev.physchem.59.032607.093602} {\bibfield
  {journal} {\bibinfo  {journal} {Annu. Rev. Phys. Chem.}\ }\textbf {\bibinfo
  {volume} {59}},\ \bibinfo {pages} {433--462} (\bibinfo {year}
  {2008})}\BibitemShut {NoStop}%
\bibitem [{\citenamefont {Gilbert}, \citenamefont {Besley},\ and\ \citenamefont
  {Gill}(2008)}]{Gilbert2008}%
  \BibitemOpen
  \bibfield  {author} {\bibinfo {author} {\bibfnamefont {A.~T.~B.}\
  \bibnamefont {Gilbert}}, \bibinfo {author} {\bibfnamefont {N.~A.}\
  \bibnamefont {Besley}}, \ and\ \bibinfo {author} {\bibfnamefont {P.~M.~W.}\
  \bibnamefont {Gill}},\ }\bibfield  {title} {\enquote {\bibinfo {title}
  {{Self-Consistent Field Calculations of Excited States Using the Maximum
  Overlap Method}},}\ }\href@noop {} {\bibfield  {journal} {\bibinfo  {journal}
  {J. Phys. Chem. A}\ }\textbf {\bibinfo {volume} {112(50)}},\ \bibinfo {pages}
  {13164--13171} (\bibinfo {year} {2008})}\BibitemShut {NoStop}%
\bibitem [{\citenamefont {Besley}, \citenamefont {Gilbert},\ and\ \citenamefont
  {Gill}(2009)}]{Besley2009}%
  \BibitemOpen
  \bibfield  {author} {\bibinfo {author} {\bibfnamefont {N.~A.}\ \bibnamefont
  {Besley}}, \bibinfo {author} {\bibfnamefont {A.~T.~B.}\ \bibnamefont
  {Gilbert}}, \ and\ \bibinfo {author} {\bibfnamefont {P.~M.~W.}\ \bibnamefont
  {Gill}},\ }\href@noop {} {\bibfield  {journal} {\bibinfo  {journal} {J. Chem.
  Phys.}\ }\textbf {\bibinfo {volume} {130}},\ \bibinfo {pages} {124308}
  (\bibinfo {year} {2009})}\BibitemShut {NoStop}%
\bibitem [{\citenamefont {Hait}\ and\ \citenamefont
  {Head-Gordon}(2020)}]{Dip:2020:roks_core}%
  \BibitemOpen
  \bibfield  {author} {\bibinfo {author} {\bibfnamefont {D.}~\bibnamefont
  {Hait}}\ and\ \bibinfo {author} {\bibfnamefont {M.}~\bibnamefont
  {Head-Gordon}},\ }\bibfield  {title} {\enquote {\bibinfo {title} {Highly
  accurate prediction of core spectra of molecules at density functional theory
  cost: Attaining sub-electronvolt error from a restricted open-shell
  kohn–sham approach},}\ }\href {\doibase 10.1021/acs.jpclett.9b03661}
  {\bibfield  {journal} {\bibinfo  {journal} {J. Phys. Chem. Lett.}\ }\textbf
  {\bibinfo {volume} {11}},\ \bibinfo {pages} {775--786} (\bibinfo {year}
  {2020})},\ \bibinfo {note} {pMID: 31917579},\ \Eprint
  {http://arxiv.org/abs/https://doi.org/10.1021/acs.jpclett.9b03661}
  {https://doi.org/10.1021/acs.jpclett.9b03661} \BibitemShut {NoStop}%
\bibitem [{\citenamefont {Cederbaum}, \citenamefont {Domcke},\ and\
  \citenamefont {Schirmer}(1980)}]{cederbaum1980cvs}%
  \BibitemOpen
  \bibfield  {author} {\bibinfo {author} {\bibfnamefont {L.~S.}\ \bibnamefont
  {Cederbaum}}, \bibinfo {author} {\bibfnamefont {W.}~\bibnamefont {Domcke}}, \
  and\ \bibinfo {author} {\bibfnamefont {J.}~\bibnamefont {Schirmer}},\
  }\bibfield  {title} {\enquote {\bibinfo {title} {Many-body theory of core
  holes},}\ }\href@noop {} {\bibfield  {journal} {\bibinfo  {journal} {Phys.
  Rev. A}\ }\textbf {\bibinfo {volume} {22}},\ \bibinfo {pages} {206} (\bibinfo
  {year} {1980})}\BibitemShut {NoStop}%
\bibitem [{\citenamefont {Peng}, \citenamefont {Copan},\ and\ \citenamefont
  {Sokolov}(2019)}]{peng2019cvsodc}%
  \BibitemOpen
  \bibfield  {author} {\bibinfo {author} {\bibfnamefont {R.}~\bibnamefont
  {Peng}}, \bibinfo {author} {\bibfnamefont {A.~V.}\ \bibnamefont {Copan}}, \
  and\ \bibinfo {author} {\bibfnamefont {A.~Y.}\ \bibnamefont {Sokolov}},\
  }\bibfield  {title} {\enquote {\bibinfo {title} {Simulating x-ray absorption
  spectra with linear-response density cumulant theory},}\ }\href@noop {}
  {\bibfield  {journal} {\bibinfo  {journal} {The Journal of Physical Chemistry
  A}\ }\textbf {\bibinfo {volume} {123}},\ \bibinfo {pages} {1840--1850}
  (\bibinfo {year} {2019})}\BibitemShut {NoStop}%
\bibitem [{\citenamefont {Wenzel}\ \emph
  {et~al.}(2015{\natexlab{b}})\citenamefont {Wenzel}, \citenamefont {Holzer},
  \citenamefont {Wormit},\ and\ \citenamefont
  {Dreuw}}]{wenzel2015CVSADCAnalysis}%
  \BibitemOpen
  \bibfield  {author} {\bibinfo {author} {\bibfnamefont {J.}~\bibnamefont
  {Wenzel}}, \bibinfo {author} {\bibfnamefont {A.}~\bibnamefont {Holzer}},
  \bibinfo {author} {\bibfnamefont {M.}~\bibnamefont {Wormit}}, \ and\ \bibinfo
  {author} {\bibfnamefont {A.}~\bibnamefont {Dreuw}},\ }\bibfield  {title}
  {\enquote {\bibinfo {title} {Analysis and comparison of cvs-adc approaches up
  to third order for the calculation of core-excited states},}\ }\href@noop {}
  {\bibfield  {journal} {\bibinfo  {journal} {The Journal of Chemical Physics}\
  }\textbf {\bibinfo {volume} {142}},\ \bibinfo {pages} {214104} (\bibinfo
  {year} {2015}{\natexlab{b}})}\BibitemShut {NoStop}%
\bibitem [{\citenamefont {Barth}\ and\ \citenamefont
  {Cederbaum}(1981)}]{cederbaum1981CVSexcitations}%
  \BibitemOpen
  \bibfield  {author} {\bibinfo {author} {\bibfnamefont {A.}~\bibnamefont
  {Barth}}\ and\ \bibinfo {author} {\bibfnamefont {L.}~\bibnamefont
  {Cederbaum}},\ }\bibfield  {title} {\enquote {\bibinfo {title} {Many-body
  theory of core-valence excitations},}\ }\href@noop {} {\bibfield  {journal}
  {\bibinfo  {journal} {Physical Review A}\ }\textbf {\bibinfo {volume} {23}},\
  \bibinfo {pages} {1038} (\bibinfo {year} {1981})}\BibitemShut {NoStop}%
\bibitem [{\citenamefont {Ågren}\ \emph {et~al.}(1997)\citenamefont {Ågren},
  \citenamefont {Carravetta}, \citenamefont {Vahtras},\ and\ \citenamefont
  {Pettersson}}]{STEX2}%
  \BibitemOpen
  \bibfield  {author} {\bibinfo {author} {\bibfnamefont {H.}~\bibnamefont
  {Ågren}}, \bibinfo {author} {\bibfnamefont {V.}~\bibnamefont {Carravetta}},
  \bibinfo {author} {\bibfnamefont {O.}~\bibnamefont {Vahtras}}, \ and\
  \bibinfo {author} {\bibfnamefont {L.~G.}\ \bibnamefont {Pettersson}},\
  }\href@noop {} {\bibfield  {journal} {\bibinfo  {journal} {Direct SCF Direct
  Static-Exchange Calculations of Electronic Spectra. Theor. Chem. Acc.}\
  }\textbf {\bibinfo {volume} {97}},\ \bibinfo {pages} {14} (\bibinfo {year}
  {1997})}\BibitemShut {NoStop}%
\bibitem [{\citenamefont {Herbst}\ and\ \citenamefont
  {Fransson}(2020)}]{herbst2020quantifying}%
  \BibitemOpen
  \bibfield  {author} {\bibinfo {author} {\bibfnamefont {M.~F.}\ \bibnamefont
  {Herbst}}\ and\ \bibinfo {author} {\bibfnamefont {T.}~\bibnamefont
  {Fransson}},\ }\bibfield  {title} {\enquote {\bibinfo {title} {Quantifying
  the error of the core-valence separation approximation},}\ }\href@noop {}
  {\bibfield  {journal} {\bibinfo  {journal} {arXiv preprint arXiv:2005.05848}\
  } (\bibinfo {year} {2020})}\BibitemShut {NoStop}%
\bibitem [{\citenamefont {Shea}\ and\ \citenamefont
  {Neuscamman}(2018)}]{Shea2018}%
  \BibitemOpen
  \bibfield  {author} {\bibinfo {author} {\bibfnamefont {J.~A.~R.}\
  \bibnamefont {Shea}}\ and\ \bibinfo {author} {\bibfnamefont {E.}~\bibnamefont
  {Neuscamman}},\ }\bibfield  {title} {\enquote {\bibinfo {title}
  {{Communication: A mean field platform for excited state quantum
  chemistry}},}\ }\href@noop {} {\bibfield  {journal} {\bibinfo  {journal} {J.
  Chem. Phys.}\ }\textbf {\bibinfo {volume} {149}},\ \bibinfo {pages} {081101}
  (\bibinfo {year} {2018})}\BibitemShut {NoStop}%
\bibitem [{\citenamefont {Shea}, \citenamefont {Gwin},\ and\ \citenamefont
  {Neuscamman}(2020)}]{Shea2020gvp}%
  \BibitemOpen
  \bibfield  {author} {\bibinfo {author} {\bibfnamefont {J.~A.~R.}\
  \bibnamefont {Shea}}, \bibinfo {author} {\bibfnamefont {E.}~\bibnamefont
  {Gwin}}, \ and\ \bibinfo {author} {\bibfnamefont {E.}~\bibnamefont
  {Neuscamman}},\ }\bibfield  {title} {\enquote {\bibinfo {title} {A
  generalized variational principle with applications to excited state mean
  field theory},}\ }\href@noop {} {\bibfield  {journal} {\bibinfo  {journal}
  {J. Chem. Theory Comput.}\ }\textbf {\bibinfo {volume} {16}},\ \bibinfo
  {pages} {1526} (\bibinfo {year} {2020})}\BibitemShut {NoStop}%
\bibitem [{\citenamefont {Zhao}\ and\ \citenamefont
  {Neuscamman}(2020{\natexlab{a}})}]{zhao2020esmf}%
  \BibitemOpen
  \bibfield  {author} {\bibinfo {author} {\bibfnamefont {L.}~\bibnamefont
  {Zhao}}\ and\ \bibinfo {author} {\bibfnamefont {E.}~\bibnamefont
  {Neuscamman}},\ }\bibfield  {title} {\enquote {\bibinfo {title} {Excited
  state mean-field theory without automatic differentiation},}\ }\href@noop {}
  {\bibfield  {journal} {\bibinfo  {journal} {J. Chem. Phys.}\ }\textbf
  {\bibinfo {volume} {152}},\ \bibinfo {pages} {204112} (\bibinfo {year}
  {2020}{\natexlab{a}})}\BibitemShut {NoStop}%
\bibitem [{\citenamefont {Hardikar}\ and\ \citenamefont
  {Neuscamman}(2020)}]{hardikar2020}%
  \BibitemOpen
  \bibfield  {author} {\bibinfo {author} {\bibfnamefont {T.~S.}\ \bibnamefont
  {Hardikar}}\ and\ \bibinfo {author} {\bibfnamefont {E.}~\bibnamefont
  {Neuscamman}},\ }\bibfield  {title} {\enquote {\bibinfo {title} {A self
  consistent field formulation of excited state mean field theory},}\
  }\href@noop {} {\bibfield  {journal} {\bibinfo  {journal} {arXiv}\ ,\
  \bibinfo {pages} {2006.02363}} (\bibinfo {year} {2020})}\BibitemShut
  {NoStop}%
\bibitem [{\citenamefont {Clune}, \citenamefont {Shea},\ and\ \citenamefont
  {Neuscamman}(2020)}]{Clune2020topesmp2}%
  \BibitemOpen
  \bibfield  {author} {\bibinfo {author} {\bibfnamefont {R.}~\bibnamefont
  {Clune}}, \bibinfo {author} {\bibfnamefont {J.~A.~R.}\ \bibnamefont {Shea}},
  \ and\ \bibinfo {author} {\bibfnamefont {E.}~\bibnamefont {Neuscamman}},\
  }\bibfield  {title} {\enquote {\bibinfo {title} {An {N}$^5$-scaling
  excited-state-specific perturbation theory},}\ }\href@noop {} {\bibfield
  {journal} {\bibinfo  {journal} {arXiv}\ ,\ \bibinfo {pages} {2003.12923}}
  (\bibinfo {year} {2020})}\BibitemShut {NoStop}%
\bibitem [{\citenamefont {Jolly}\ and\ \citenamefont
  {Hendrickson}(1970)}]{jolly1970}%
  \BibitemOpen
  \bibfield  {author} {\bibinfo {author} {\bibfnamefont {W.~L.}\ \bibnamefont
  {Jolly}}\ and\ \bibinfo {author} {\bibfnamefont {D.~N.}\ \bibnamefont
  {Hendrickson}},\ }\bibfield  {title} {\enquote {\bibinfo {title}
  {Thermodynamic interpretation of chemical shifts in core-electron binding
  energies},}\ }\href@noop {} {\bibfield  {journal} {\bibinfo  {journal} {J.
  Am. Chem. Soc.}\ }\textbf {\bibinfo {volume} {92}},\ \bibinfo {pages}
  {1863--1871} (\bibinfo {year} {1970})}\BibitemShut {NoStop}%
\bibitem [{\citenamefont {{\AA}gren}\ \emph {et~al.}(1988)\citenamefont
  {{\AA}gren}, \citenamefont {Medina-Llanos}, \citenamefont {Mikkelsen},\ and\
  \citenamefont {Jensen}}]{aagren1988}%
  \BibitemOpen
  \bibfield  {author} {\bibinfo {author} {\bibfnamefont {H.}~\bibnamefont
  {{\AA}gren}}, \bibinfo {author} {\bibfnamefont {C.}~\bibnamefont
  {Medina-Llanos}}, \bibinfo {author} {\bibfnamefont {K.~V.}\ \bibnamefont
  {Mikkelsen}}, \ and\ \bibinfo {author} {\bibfnamefont {H.-J.~A.}\
  \bibnamefont {Jensen}},\ }\bibfield  {title} {\enquote {\bibinfo {title} {On
  the validity of the equivalent core approximation in born-haber analyses of
  liquids and solutions},}\ }\href@noop {} {\bibfield  {journal} {\bibinfo
  {journal} {Chem. Phys. Lett.}\ }\textbf {\bibinfo {volume} {153}},\ \bibinfo
  {pages} {322--327} (\bibinfo {year} {1988})}\BibitemShut {NoStop}%
\bibitem [{\citenamefont {Takahashi}(2017)}]{RelativisticCorrections}%
  \BibitemOpen
  \bibfield  {author} {\bibinfo {author} {\bibfnamefont {O.}~\bibnamefont
  {Takahashi}},\ }\bibfield  {title} {\enquote {\bibinfo {title} {Relativistic
  corrections for single-and double-core excitation at the k-and l-edges from
  li to kr},}\ }\href@noop {} {\bibfield  {journal} {\bibinfo  {journal}
  {Computational and Theoretical Chemistry}\ }\textbf {\bibinfo {volume}
  {1102}},\ \bibinfo {pages} {80--86} (\bibinfo {year} {2017})}\BibitemShut
  {NoStop}%
\bibitem [{\citenamefont {Schirmer}\ \emph
  {et~al.}(1993{\natexlab{a}})\citenamefont {Schirmer}, \citenamefont
  {Trofimov}, \citenamefont {Randall}, \citenamefont {Feldhaus}, \citenamefont
  {Bradshaw}, \citenamefont {Ma}, \citenamefont {Chen},\ and\ \citenamefont
  {Sette}}]{WaterAmmoniaMethane}%
  \BibitemOpen
  \bibfield  {author} {\bibinfo {author} {\bibfnamefont {J.}~\bibnamefont
  {Schirmer}}, \bibinfo {author} {\bibfnamefont {A.}~\bibnamefont {Trofimov}},
  \bibinfo {author} {\bibfnamefont {K.}~\bibnamefont {Randall}}, \bibinfo
  {author} {\bibfnamefont {J.}~\bibnamefont {Feldhaus}}, \bibinfo {author}
  {\bibfnamefont {A.}~\bibnamefont {Bradshaw}}, \bibinfo {author}
  {\bibfnamefont {Y.}~\bibnamefont {Ma}}, \bibinfo {author} {\bibfnamefont
  {C.}~\bibnamefont {Chen}}, \ and\ \bibinfo {author} {\bibfnamefont
  {F.}~\bibnamefont {Sette}},\ }\bibfield  {title} {\enquote {\bibinfo {title}
  {K-shell excitation of the water, ammonia, and methane molecules using
  high-resolution photoabsorption spectroscopy},}\ }\href@noop {} {\bibfield
  {journal} {\bibinfo  {journal} {Physical Review A}\ }\textbf {\bibinfo
  {volume} {47}},\ \bibinfo {pages} {1136} (\bibinfo {year}
  {1993}{\natexlab{a}})}\BibitemShut {NoStop}%
\bibitem [{\citenamefont {Tronc}, \citenamefont {King},\ and\ \citenamefont
  {Read}(1979)}]{tronc1979carbon}%
  \BibitemOpen
  \bibfield  {author} {\bibinfo {author} {\bibfnamefont {M.}~\bibnamefont
  {Tronc}}, \bibinfo {author} {\bibfnamefont {G.~C.}\ \bibnamefont {King}}, \
  and\ \bibinfo {author} {\bibfnamefont {F.}~\bibnamefont {Read}},\ }\bibfield
  {title} {\enquote {\bibinfo {title} {Carbon k-shell excitation in small
  molecules by high-resolution electron impact},}\ }\href@noop {} {\bibfield
  {journal} {\bibinfo  {journal} {Journal of Physics B: Atomic and Molecular
  Physics}\ }\textbf {\bibinfo {volume} {12}},\ \bibinfo {pages} {137}
  (\bibinfo {year} {1979})}\BibitemShut {NoStop}%
\bibitem [{\citenamefont {Hitchcock}\ and\ \citenamefont
  {Brion}(1977)}]{hitchcock1977carbon}%
  \BibitemOpen
  \bibfield  {author} {\bibinfo {author} {\bibfnamefont {A.}~\bibnamefont
  {Hitchcock}}\ and\ \bibinfo {author} {\bibfnamefont {C.}~\bibnamefont
  {Brion}},\ }\bibfield  {title} {\enquote {\bibinfo {title} {Carbon k-shell
  excitation of c2h2, c2h4, c2h6 and c6h6 by 2.5 kev electron impact},}\
  }\href@noop {} {\bibfield  {journal} {\bibinfo  {journal} {Journal of
  Electron Spectroscopy and Related Phenomena}\ }\textbf {\bibinfo {volume}
  {10}},\ \bibinfo {pages} {317--330} (\bibinfo {year} {1977})}\BibitemShut
  {NoStop}%
\bibitem [{\citenamefont {Hitchcock}\ and\ \citenamefont
  {Brion}(1979)}]{hitchcock1979inner}%
  \BibitemOpen
  \bibfield  {author} {\bibinfo {author} {\bibfnamefont {A.}~\bibnamefont
  {Hitchcock}}\ and\ \bibinfo {author} {\bibfnamefont {C.}~\bibnamefont
  {Brion}},\ }\bibfield  {title} {\enquote {\bibinfo {title} {Inner shell
  electron energy loss studies of hcn and c2n2},}\ }\href@noop {} {\bibfield
  {journal} {\bibinfo  {journal} {Chemical Physics}\ }\textbf {\bibinfo
  {volume} {37}},\ \bibinfo {pages} {319--331} (\bibinfo {year}
  {1979})}\BibitemShut {NoStop}%
\bibitem [{\citenamefont {Remmers}\ \emph {et~al.}(1992)\citenamefont
  {Remmers}, \citenamefont {Domke}, \citenamefont {Puschmann}, \citenamefont
  {Mandel}, \citenamefont {Xue}, \citenamefont {Kaindl}, \citenamefont
  {Hudson},\ and\ \citenamefont {Shirley}}]{remmers1992high}%
  \BibitemOpen
  \bibfield  {author} {\bibinfo {author} {\bibfnamefont {G.}~\bibnamefont
  {Remmers}}, \bibinfo {author} {\bibfnamefont {M.}~\bibnamefont {Domke}},
  \bibinfo {author} {\bibfnamefont {A.}~\bibnamefont {Puschmann}}, \bibinfo
  {author} {\bibfnamefont {T.}~\bibnamefont {Mandel}}, \bibinfo {author}
  {\bibfnamefont {C.}~\bibnamefont {Xue}}, \bibinfo {author} {\bibfnamefont
  {G.}~\bibnamefont {Kaindl}}, \bibinfo {author} {\bibfnamefont
  {E.}~\bibnamefont {Hudson}}, \ and\ \bibinfo {author} {\bibfnamefont
  {D.}~\bibnamefont {Shirley}},\ }\bibfield  {title} {\enquote {\bibinfo
  {title} {High-resolution k-shell photoabsorption in formaldehyde},}\
  }\href@noop {} {\bibfield  {journal} {\bibinfo  {journal} {Physical Review
  A}\ }\textbf {\bibinfo {volume} {46}},\ \bibinfo {pages} {3935} (\bibinfo
  {year} {1992})}\BibitemShut {NoStop}%
\bibitem [{\citenamefont {Hempelmann}\ \emph {et~al.}(1999)\citenamefont
  {Hempelmann}, \citenamefont {Piancastelli}, \citenamefont {Heiser},
  \citenamefont {Gessner}, \citenamefont {R{\"u}del},\ and\ \citenamefont
  {Becker}}]{hempelmann1999resonant}%
  \BibitemOpen
  \bibfield  {author} {\bibinfo {author} {\bibfnamefont {A.}~\bibnamefont
  {Hempelmann}}, \bibinfo {author} {\bibfnamefont {M.}~\bibnamefont
  {Piancastelli}}, \bibinfo {author} {\bibfnamefont {F.}~\bibnamefont
  {Heiser}}, \bibinfo {author} {\bibfnamefont {O.}~\bibnamefont {Gessner}},
  \bibinfo {author} {\bibfnamefont {A.}~\bibnamefont {R{\"u}del}}, \ and\
  \bibinfo {author} {\bibfnamefont {U.}~\bibnamefont {Becker}},\ }\bibfield
  {title} {\enquote {\bibinfo {title} {Resonant photofragmentation of methanol
  at the carbon and oxygen k-edge by high-resolution ion-yield spectroscopy},}\
  }\href@noop {} {\bibfield  {journal} {\bibinfo  {journal} {Journal of Physics
  B: Atomic, Molecular and Optical Physics}\ }\textbf {\bibinfo {volume}
  {32}},\ \bibinfo {pages} {2677} (\bibinfo {year} {1999})}\BibitemShut
  {NoStop}%
\bibitem [{\citenamefont {Tronc}, \citenamefont {King},\ and\ \citenamefont
  {Read}(1980)}]{tronc1980nitrogen}%
  \BibitemOpen
  \bibfield  {author} {\bibinfo {author} {\bibfnamefont {M.}~\bibnamefont
  {Tronc}}, \bibinfo {author} {\bibfnamefont {G.~C.}\ \bibnamefont {King}}, \
  and\ \bibinfo {author} {\bibfnamefont {F.}~\bibnamefont {Read}},\ }\bibfield
  {title} {\enquote {\bibinfo {title} {Nitrogen k-shell excitation in n2, no
  and n2o by high-resolution electron energy-loss spectroscopy},}\ }\href@noop
  {} {\bibfield  {journal} {\bibinfo  {journal} {Journal of Physics B: Atomic
  and Molecular Physics}\ }\textbf {\bibinfo {volume} {13}},\ \bibinfo {pages}
  {999} (\bibinfo {year} {1980})}\BibitemShut {NoStop}%
\bibitem [{\citenamefont {Barrus}\ \emph {et~al.}(1979)\citenamefont {Barrus},
  \citenamefont {Blake}, \citenamefont {Burek}, \citenamefont {Chambers},\ and\
  \citenamefont {Pregenzer}}]{Barrus1979OKedge}%
  \BibitemOpen
  \bibfield  {author} {\bibinfo {author} {\bibfnamefont {D.~M.}\ \bibnamefont
  {Barrus}}, \bibinfo {author} {\bibfnamefont {R.~L.}\ \bibnamefont {Blake}},
  \bibinfo {author} {\bibfnamefont {A.~J.}\ \bibnamefont {Burek}}, \bibinfo
  {author} {\bibfnamefont {K.~C.}\ \bibnamefont {Chambers}}, \ and\ \bibinfo
  {author} {\bibfnamefont {A.~L.}\ \bibnamefont {Pregenzer}},\ }\bibfield
  {title} {\enquote {\bibinfo {title} {$k$-shell photoabsorption coefficients
  of ${\mathrm{o}}_{2}$, c${\mathrm{o}}_{2}$, co, and ${\mathrm{n}}_{2}$o},}\
  }\href {\doibase 10.1103/PhysRevA.20.1045} {\bibfield  {journal} {\bibinfo
  {journal} {Phys. Rev. A}\ }\textbf {\bibinfo {volume} {20}},\ \bibinfo
  {pages} {1045--1061} (\bibinfo {year} {1979})}\BibitemShut {NoStop}%
\bibitem [{\citenamefont {Hitchcock}\ and\ \citenamefont
  {Brion}(1981)}]{hitchcock1981F}%
  \BibitemOpen
  \bibfield  {author} {\bibinfo {author} {\bibfnamefont {A.}~\bibnamefont
  {Hitchcock}}\ and\ \bibinfo {author} {\bibfnamefont {C.}~\bibnamefont
  {Brion}},\ }\bibfield  {title} {\enquote {\bibinfo {title} {K-shell
  excitation of hf and f2 studied by electron energy-loss spectroscopy},}\
  }\href@noop {} {\bibfield  {journal} {\bibinfo  {journal} {Journal of Physics
  B: Atomic and Molecular Physics}\ }\textbf {\bibinfo {volume} {14}},\
  \bibinfo {pages} {4399} (\bibinfo {year} {1981})}\BibitemShut {NoStop}%
\bibitem [{\citenamefont {Schirmer}\ \emph
  {et~al.}(1993{\natexlab{b}})\citenamefont {Schirmer}, \citenamefont
  {Trofimov}, \citenamefont {Randall}, \citenamefont {Feldhaus}, \citenamefont
  {Bradshaw}, \citenamefont {Ma}, \citenamefont {Chen},\ and\ \citenamefont
  {Sette}}]{schirmer1993}%
  \BibitemOpen
  \bibfield  {author} {\bibinfo {author} {\bibfnamefont {J.}~\bibnamefont
  {Schirmer}}, \bibinfo {author} {\bibfnamefont {A.~B.}\ \bibnamefont
  {Trofimov}}, \bibinfo {author} {\bibfnamefont {K.~J.}\ \bibnamefont
  {Randall}}, \bibinfo {author} {\bibfnamefont {J.}~\bibnamefont {Feldhaus}},
  \bibinfo {author} {\bibfnamefont {A.~M.}\ \bibnamefont {Bradshaw}}, \bibinfo
  {author} {\bibfnamefont {Y.}~\bibnamefont {Ma}}, \bibinfo {author}
  {\bibfnamefont {C.~T.}\ \bibnamefont {Chen}}, \ and\ \bibinfo {author}
  {\bibfnamefont {F.}~\bibnamefont {Sette}},\ }\bibfield  {title} {\enquote
  {\bibinfo {title} {K-shell excitation of the water, ammonia, and methane
  molecules using high-resolution photoabsorption spectroscopy},}\ }\href@noop
  {} {\bibfield  {journal} {\bibinfo  {journal} {Phys. Rev. A}\ }\textbf
  {\bibinfo {volume} {47}},\ \bibinfo {pages} {1136} (\bibinfo {year}
  {1993}{\natexlab{b}})}\BibitemShut {NoStop}%
\bibitem [{\citenamefont {Zhao}\ and\ \citenamefont
  {Neuscamman}(2020{\natexlab{b}})}]{Zhao2019dft}%
  \BibitemOpen
  \bibfield  {author} {\bibinfo {author} {\bibfnamefont {L.}~\bibnamefont
  {Zhao}}\ and\ \bibinfo {author} {\bibfnamefont {E.}~\bibnamefont
  {Neuscamman}},\ }\bibfield  {title} {\enquote {\bibinfo {title} {Density
  functional extension to excited-state mean-field theory},}\ }\href@noop {}
  {\bibfield  {journal} {\bibinfo  {journal} {J. Chem. Theory Comput.}\
  }\textbf {\bibinfo {volume} {16}},\ \bibinfo {pages} {164} (\bibinfo {year}
  {2020}{\natexlab{b}})}\BibitemShut {NoStop}%
\end{thebibliography}%

\clearpage

\end{document}